\documentclass[12pt]{article}
\usepackage{amsfonts}
\usepackage{latexsym,amsmath,amssymb}
\usepackage{graphics,epstopdf}
\usepackage[pdftex]{graphicx}
\usepackage {tikz}
\usetikzlibrary {positioning}
\definecolor {processblue}{cmyk}{0.96,0,0,0}
\numberwithin{equation}{section}

\newtheorem{thm}{Theorem}[section]
\newtheorem{cor}[thm]{Corollary}
\newtheorem{lem}[thm]{Lemma}
\newtheorem{prop}[thm]{Proposition}

\numberwithin{equation}{section}

\begin{document}

\bigskip

\bigskip

\begin{center}
{\Large \textbf{Algorithms and Identities for B$\acute{e}$zier curves via Post Quantum Blossom}}

\bigskip

\textbf{Alaa Mohammed Obad}$^{1}~~$\textbf{Khalid Khan}$^{2}~~$\textbf{D.K. Lobiyal}$^{3}~$and \textbf{Asif Khan}$^{4}$\\
$^{1,4}$Department of Mathematics, Aligarh Muslim University, Aligarh-202002, India\\

$^{2,3}$School of Computer and System Sciences, SC \& SS, J.N.U., New Delhi-110067, India%

allaobad4@gmail.com; khalidga1517@gmail.com; dklobiyal@gmail.com; akhan.mm@amu.ac.in

\bigskip

\bigskip

\textbf{Abstract}
\end{center}
In this paper, a new analogue of blossom based on post quantum calculus is introduced. The post quantum blossom has been adapted for developing identities and algorithms for Bernstein bases and B$\acute{e}$zier curves. By applying the post quantum blossom, various new identities and formulae expressing the monomials in terms of the post quantun Bernstein
basis functions and a post quantun variant of Marsden's identity are investigated. For each post quantum B$\acute{e}$zier curves of degree $m,$ a collection of $m!$ new, affine invariant, recursive evaluation algorithms are derived.

\bigskip

{\footnotesize \emph{Keywords and phrases}: post quantum integers; post quantum blossom; de Casteljau algorithm; Marsden's identity; post quantum Bernstein polynomials; post quantum B$\acute{e}$zier curve; quantum Bernstein polynomials.\\

{\footnotesize \emph{MSC: primary 65D17; secondary 41A10, 41A25, 41A36.}: \newline

\bigskip

\section{Introduction}

Approximation theory basically deals with approximation of functions by simpler functions or more easily calculated functions. Broadly it is divided into theoretical and constructive approximation \cite{ ss8,ss17,pp,mhaskar}.\\

Mursaleen et al applied post quantum calculus in constructive approximation theory and introduced the first post quantum analogue of Bernstein operators \cite{mka1} based on post quantum integers.
The post quantum Bernstein operators introduced by them is generalization of well known classical Bernstein \cite{brn} operators and Phillips quantum Bernstein operators (polynomials) \cite{hp,pl}. For literatures related to constructive approximation theory, quantum calculus and post quantum calculus, one can see \cite{ss6,kadak2,ss13,ss17,ss18,lp,ss21,ss22,ss24,cogent,ss25,m4,zmn,mnak1,ma1,ss23,ss26,mnaa,nasir1,mah,ss29,ss30,vp,ss31,wei}.\\

Similarly Computer aided geometric design (CAGD) is a discipline which deals with computational aspects of geometric objects. If emphasizes on the mathematical development of curves and surfaces such that it become compatible with computers. The kernel of CAGD lies in approximation theory and numerical analysis. Bases of Bernstein operators have been used to draw curves and surfaces. B$\acute{e}$zier curves were independently developed by P.de Casteljan at Citroen and by P. B$\acute{e}$zier at Renault. For details on Bezier curves and surfaces approximation, one can refer \cite{bezier, farouki, goldman,wcq,jag, khalid3,khalidthesis,rababah,thomas}.\\

 Khan and Lobiyal \cite{khalid2} recently constructed post quantum analogue of Lupa\c{s} quantum Bernstein operators (rational) and investigated various properties of Lupa\c{s}  post quantum B$\acute{e}$zier curves and Surfaces. For some applications of the extra parameter `$p$' of post quantum analogue in terms of flexibility to design geometric shapes and for flexibility in approximation, one can refer \cite{khalid2,m4}.\\
 
  In CAGD, blossoming method deals with representation of curves into simpler form like representing a polynomial of degree $m$ into monomial in $m$ variables each of degree one.  Blossoming method is used to reduce computational complexity for construction of B$\acute{e}$zier curves and surfaces. This provides a powerful tool for deriving identities and developing change of basis algorithms for basis and B$\acute{e}$zier curves. In \cite{Simeonova}, some algorithms and identities for quantum Bernstein bases and quantum B$\acute{e}$zier curves using the method of quantum Blossoming has been constructed.\\
  
   Motivated by above mentioned work, the idea of post quantum calculus and its importance, in next sections, we will investigate and derive several results via post quantum analogue of blossoming. The post quantum blossom will be used for developing identities and algorithms for  Bernstein bases and B$\acute{e}$zier curves. By applying the post quantum blossom, various new identities and formulae expressing the monomials in terms of the post quantum Bernstein
basis functions and a post quantun variant of Marsden's identity will be investigated. For each post quantum B$\acute{e}$zier curves of degree $m$, a collection of $m!$ new, affine invariant, recursive evaluation algorithms will be derived.\\

Let us recall certain notations and definitions from post quantum calculus,\\

The post quantum number is defined by, for any number $m$ 
\begin{equation*}
\lbrack
m]_{p,q}=p^{m-1}+p^{m-2}q+p^{m-3}q^2+\cdots+pq^{m-2}+q^{m-1}\\
=\left\{
\begin{array}{lll}
\frac{p^{m}-q^{m}}{p-q},~~~~~~~~\mbox{when $~~p\neq q $  } & \\
&  \\
m~p^{m-1},~~~~~~\mbox{ when~~ $p=q$  }. & \\
\end{array}%
\right.
\end{equation*}

This paper has been arranged in the following way: In Section $2$ and $3,$ we introduce the basic
definitions, fundamental formulas, and explicit notation for post quantum Bernstein bases and post quantum B$\acute{e}$zier
curves. In Section $4$, we define the post quantum blossom and establish the existence and the uniqueness
of this blossom. In Section $5$, we invoke post quantum blossoming to develop novel evaluation algorithms
for post quantum B$\acute{e}$zier curves and in Section $6,$ we use the post quantum blossom to derive new identities involving
the post quantum Bernstein basis functions, including a post quantum version of Marsdens identity as well as formulas for representing monomials in terms of the post quantum Bernstein basis functions.

\section{Post quantum Bernstein basis functions}

The post quantum Bernstein basis function \cite{khalid1,mka1} is as follows
\begin{equation}\label{ee4}
B^{r,m}_{p,q}(t)=\frac{1}{p^{\frac{m(m-1)}{2}}}\left[
\begin{array}{c}
m \\
r%
\end{array}%
\right] _{p,q} p^{\frac{r(r-1)}{2}}~~t^{r}(1-t)^{m-r}_{p,q} ,~~~~~t\in \lbrack 0,1]
\end{equation}\\
where
\begin{equation*}
(1-t)^{m-r}_{p,q}=\prod\limits_{s=0}^{m-r-1}(p^s-q^{s}t).
\end{equation*}

%

 \begin{thm}\cite{khalid1}
Each post quantum Bernstein function of degree $m$ is a linear combination of two post quantum Bernstein functions of degree $m+1.$
\begin{equation}\label{e8}
B^{r,m}_{p,q}(t)= \bigg(\frac{p^{r}~{[m+1-r]}_{p,q}}{{[m+1]}_{p,q}}\bigg)~B^{r,m+1}_{p,q}(t)+\bigg(1-\frac{p^{r+1}~{[m-r]}_{p,q}}{{[m+1]}_{p,q}}\bigg)~B^{r+1,m+1}_{p,q}(t).
\end{equation}
\end{thm}

Throughout the paper onwards, we use $B_r^m(t;p,q)$ in place of $B^{r,m}_{p,q}(t).$

\section{Post quantum Bernstein B$\acute{e}$zier curves}

The post quantum B$\acute{e}$zier curves \cite{khalid1} of degree $m$ using the post quantum analogues of the Bernstein basis functions are as follows:
\begin{equation}\label{e12}
{\bf{ P}}(t; p,q) = \sum\limits_{i=0}^{m} {\bf{P_i}}~B_i^m(t;p,q)
\end{equation}

where $P_i \in R^3$  $(i = 0, 1, \cdots  , m),$ $P_i$ are control points. Joining up adjacent points $P_i,$  $i = 0, 1, 2,  \cdots , m $ to obtain a
polygon which is called the control polygon of post quantum B$\acute{e}$zier curves.

\subsection{de Casteljau algorithm}

 
 Let ${\bf{{\tilde{P}}^{0}_{i}}}(t;p,q)={\bf{{\hat{p}}^{0}_{i}}}(t;p,q)={\bf{p_i}}$~,~~~$i=0,1, \cdots , m.$ Define
\begin{equation}\label{ee2.3}
 {\bf{{\tilde{P}}^{k}_{i}}}(t;p,q)=(p^{m-k}-p^i~q^{m-k-i}~t){\bf{{\tilde{P}}^{k-1}_{i}}}(t;p,q)+p^i~q^{m-k-i}~t~{\bf{{\tilde{P}}^{k-1}_{i+1}}}(t;p,q)
\end{equation}
and
\begin{equation}\label{ee2.4}
{\bf{{\hat{p}}^{k}_{i}}}(t;p,q)=q^i(p^{m-k-i}-q^{m-k-i}~t){\bf{{\hat{p}}^{k-1}_{i}}}(t;p,q)+p^{m-k}~t~{\bf{{\hat{p}}^{k-1}_{i+1}}}(t;p,q)
\end{equation}
\\
$i=0,1, \cdots , m-k ,~~~ k=1, \cdots , m.$ 
~Then ~ ${\bf{{\tilde{P}}^{m}_{0}}}(t;p,q)={\bf{{\hat{p}}^{m}_{0}}}(t;p,q)={\bf{p(t;p,q)}}.$\\

Now for illustration purpose, we present Figure 1 and 2 for cubic post quantum B$\acute{e}$zier curves using the above two de Casteljau algorithms. In both algorithms, the property of affine invariant holds at only for the final node at the top of diagram. However for p=1, the property of affine invariant holds at every intermediate nodes in first algorithms, in second algorithm this property holds only for the final node at the top of diagram.\\



P. Simeonova et al. \cite{Simeonova} has given a  new approach to identities and algorithms for quantum Bernstein bases and quantum B$\acute{e}$zier curves using quantum blossom. In this paper we extend these results for post quantum Bernstein bases and post quantum B$\acute{e}$zier curves using post quantum blossoming.\\

Four main contributions of this paper are:\\

\textbf{Blossoming:}  The post quantum blossom, a new variant of the blossom has been introduced which will prove new identities for post quantum Bernstein bases and generate new approach for post quantum B$\acute{e}$zier curves.\\

\textbf{Identities:} Using post quantum blossom, new identities are derived for the post quantum Bernstein bases, and a post quantum variant of Marsden's identity and monomials get represented using an explicit formula in terms of the post quantum Bernstein basis functions.\\

\textbf{Recursive Evaluation Algorithms:} Using post quantum blossom technique, for a given post quantum B$\acute{e}$zier curve of degree $m$, $m!$ new affine invariant, recursive evaluation algorithms has been constructed.\\



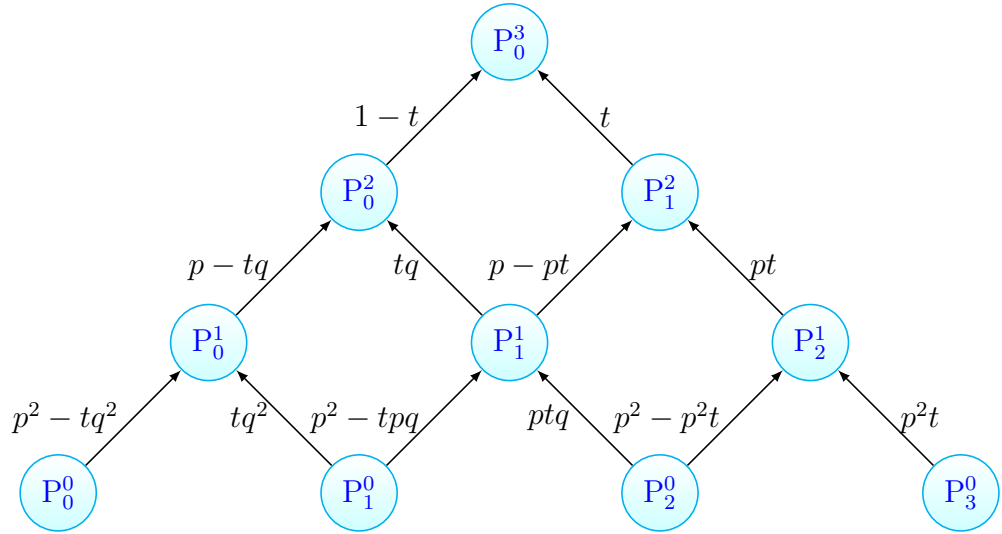
\begin{figure*}[htb!]
	\begin {center}
	\begin {tikzpicture}[-latex ,auto ,node distance =2 cm and 2cm ,on grid ,
	semithick ,
	state/.style ={ circle ,top color =white , bottom color = processblue!20 ,
		draw,processblue , text=blue , minimum width =.5 cm}]
	\node[state] (A)
	{\~{p}$_0^3$};
	\node[state] (B) [below left=of A] {\~{P}$_0^2$};
	\node[state] (C) [below right =of A] {\~{P}$_1^2$};
	\node[state] (D) [below left=of B] {\~{P}$_0^1$};
	\node[state] (E) [below right=of B] {\~{P}$_1^1$};
	\node[state] (F) [below right=of C] {\~{P}$_2^1$};
	\node[state] (G) [below left=of D] {\~{P}$_0^0$};
	\node[state] (H) [below right=of D] {\~{P}$_1^0$};
	\node[state] (I) [below right=of E] {\~{P}$_2^0$};
	\node[state] (J) [below right=of F] {\~{P}$_3^0$};
	
	\path (B) edge [bend right =0] node[left = 0.05 cm] {$1-t$} (A);
	\path (C) edge [bend right = 0] node[right =0.05 cm] {$t$} (A);
	\path (D) edge [bend left =0] node[left =0.05 cm] {$p-tq$} (B);
	\path (E) edge [bend left =0] node[left =0.05 cm] {$tq$} (B);
	\path (E) edge [bend left =0] node[left =0.05 cm] {$p-pt$} (C);
	\path (F) edge [bend right = 0] node[right =0.05 cm] {$pt$} (C);
	\path (G) edge [bend right = 0] node[left =0.05 cm] {$p^{2}-tq^2$} (D);
	\path (H) edge [bend right = 0] node[left =0.05 cm] {$tq^2$} (D);
	\path (H) edge [bend right = 0] node[left =0.05 cm] {$p^{2}-tpq$} (E);
	\path (I) edge [bend right = 0] node[left =0.05 cm] {$ptq$} (E);
	\path (I) edge [bend right = 0] node[left =0.05 cm] {$p^2-p^2t$} (F);
	\path (J) edge [bend right = 0] node[right =0.05 cm] {$p^2t$} (F);
	
\end{tikzpicture}
\end{center}
\caption{`The first de-Casteljau evaluation algorithm for a cubic post quantum B$\acute{e}$zier curve on the interval $[0,1]$'}\label{f1}
\end{figure*}

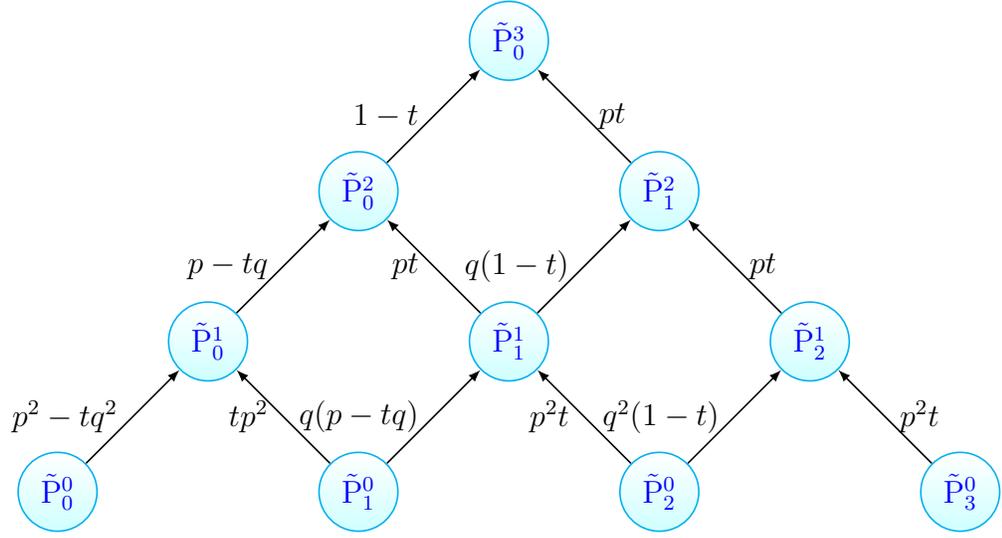
\begin{figure*}[htb!]
\begin {center}
\begin {tikzpicture}[-latex ,auto ,node distance =2 cm and 2cm ,on grid ,
semithick ,
state/.style ={ circle ,top color =white , bottom color = processblue!20 ,
draw,processblue , text=blue , minimum width =.5 cm}]
\node[state] (A)
{\^{P}$_0^3$};
\node[state] (B) [below left=of A] {\^{P}$_0^2$};
\node[state] (C) [below right =of A] {\^{P}$_1^2$};
\node[state] (D) [below left=of B] {\^{P}$_0^1$};
\node[state] (E) [below right=of B] {\^{P}$_1^1$};
\node[state] (F) [below right=of C] {\^{P}$_2^1$};
\node[state] (G) [below left=of D] {\^{P}$_0^0$};
\node[state] (H) [below right=of D] {\^{P}$_1^0$};
\node[state] (I) [below right=of E] {\^{P}$_2^0$};
\node[state] (J) [below right=of F] {\^{P}$_3^0$};

\path (B) edge [bend right =0] node[left = 0.05 cm] {$1-t$} (A);
\path (C) edge [bend right = 0] node[right =0.05 cm] {$t$} (A);
\path (D) edge [bend left =0] node[left =0.05 cm] {$p-tq$} (B);
\path (E) edge [bend left =0] node[left =0.05 cm] {$pt$} (B);
\path (E) edge [bend left =0] node[left =0.05 cm] {$q(1-t)$} (C);
\path (F) edge [bend right = 0] node[right =0.05 cm] {$pt$} (C);
\path (G) edge [bend right = 0] node[left =0.05 cm] {$p^2-tq^2$} (D);
\path (H) edge [bend right = 0] node[left =0.05 cm] {$tp^2$} (D);
\path (H) edge [bend right = 0] node[left =0.05 cm] {$q(p-tq)$} (E);
\path (I) edge [bend right = 0] node[left =0.05 cm] {$p^2t$} (E);
\path (I) edge [bend right = 0] node[left =0.05 cm] {$q^2(1-t)$} (F);
\path (J) edge [bend right = 0] node[right =0.05 cm] {$p^2t$} (F);

\end{tikzpicture}
\end{center}
\caption{`The second de-Casteljau evaluation algorithm for a cubic post quantum B$\acute{e}$zier curve on the interval $[0,1]$'}\label{f2}
\end{figure*}

\section{Post quantum blossoming}

Blossoming has given new approach for deriving identities and developing change of basis algorithms for standard Bernstein bases and B$\acute{e}$zier curves \cite{Simeonova}. In this section, post quantum blossoming as an extension of standard quantum blossoming has been achieved.\\

The post quantum blossom or post quantum polar form of a polynomial $S(t)$ of degree $m$ is the unique symmetric multiaffine function $s(u_1, \cdots , u_m; p, q)$ that reduces to $S(t)$ along the post quantum diagonal. That is, $s(u_1, \cdots , u_m; p, q)$ is the unique multivariate polynomial satisfying the following three axioms:
\\

\textbf{Post quantum Blossoming axioms}\\

1. \textbf{Symmetry:}
$s(u_1, \cdots , u_m; p, q)$  =$s(u_{\sigma(1)}, \cdots , u_{\sigma(m)}; p,q)$
for every permutation $\sigma$ of the set $\{1,2, \cdots , m\}.$\\

2. \textbf{Multiaffine:}

$s(u_1, . . . , (1 - \alpha )u_k + \alpha v_k , . . . , u_m; p,q) = (1 - \alpha)~p(u_1, . . . , u_k , . . . , u_m;p, q)
+\alpha ~ s(u_1, . . . , v_k , . . . , u_m;p, q)$\\

3. \textbf{Post quantum Diagonal:}

$s(p^{m-1}t, p^{m-2}tq, \cdots, tq^{m-1};p, q) = S(t).$\\\\
The multiaffine property is equivalent to the fact that each variable $u_1, \cdots , u_m $ appears to at most the first power that is,  $s(u_1, \cdots , u_m; p, q)$ is a polynomial of degree at most one in each variable.\\ post quantum blossoming gets interesting due to Dual functional property, which get derived and that relate post quantum blossom of a polynomial to its post quantum B$\acute{e}$zier control points.\\

\textbf{Dual functional property}\\

Let S(t) be a post quantum B$\acute{e}$zier curve of degree $m$ over the interval $[0, 1]$ with control points $P_0, \cdots , P_m$ and let $s(u_1, \cdots , u_m; p, q)$ be the post quantum blossom of S(t). Then

\begin{equation}\label{e4.1}
P_k = s(0, \cdots , 0, p^{m-1}, p^{m-2}q, \cdots , p^{m-1}q^{k-1};p, q),~~ k = 0,1, \cdots , m.
\end{equation}

This Dual functional property gets proved in Theorem \ref{t4.4}. \\

Now we establish those functions existence and uniqueness which satisfy post quantum blossoming axioms, subject to restrictions on $p,q$ for all polynomials of degree $m$. But before proceeding to it let's get feel of post quantum blossom by computing the post quantum blossom for some simple cases.\\

\textbf{Post quantum Blossom of cubic polynomials}

Let us consider a cubic polynomial represented by the monomial $1, t, t^2, \text{and}~~ t^3$. Now these monomials can be easily post quantum blossomed for any $p\neq0,$ and $q \neq0$, since in each case the associated function $s(u_1, u_2, u_3;p, q)$ given below can be easily verified as it is symmetric, multiaffine, and reduces to the required monomial along the post quantum diagonal:

$$S(t)=1\Rightarrow s(u_1,u_2,u_3;p,q)=1,$$

$$S(t)=t \Rightarrow s(u_1,u_2,u_3;p,q)=\frac{u_1+u_2+u_3}{(p^2+pq+q^2)},$$

$$S(t)=t^2 \Rightarrow s(u_1,u_2,u_3;p,q)=\frac{u_1 u_2+u_2 u_3+u_3u_1}{pq(p^2+pq+q^2)},$$

$$S(t)=t^3 \Rightarrow s(u_1,u_2,u_3;p,q)=\frac{u_1 u_2 u_3}{p^3q^3}.$$

In the right hand side of the above equation, it can be seen that functions in numerator are combinations of three variables which are written in symmetrical fashion while in case of denominator function is evaluated in symmetrical order at $p^2,~ pq$ and $ q^2.$ Using these results, any cubic polynomial $S(t) = a_3t^3 + a_2t^2 + a_1 t + a_0 $ for $ p\neq 0, q \neq 0$ can be post quantum blossom by setting

$$s(u_1,u_2,u_3;p,q)=a_3 \frac{u_1 u_2 u_3}{p^3q^3}+a_2 \frac{u_1 u_2+u_2 u_3+u_3u_1}{pq(p^2+pq+q^2)}+a_1 \frac{u_1+u_2+u_3}{(p^2+pq+q^2)} +a_0. $$

Note: For $p=q=1,$ above blossoming reduces into classical blossoming of cubic polynomials.\\

Similarly, we can apply post quantum blossom techniques for polynomials of degree $m$ by first post quantum blossoming
the monomials $t^k$ , for $k = 0, \cdots , m,$ and then applying linearity.
Indeed, let
$$ \phi_{m,k}(u_1,u_2,\cdots,u_m)=\sum\limits_{1\leq i_1 <i_2<\cdots< i_k \leq m} ~~~~u_{i_1} \cdots u_{i_k}$$
where the sum runs over all subsets $\{i_1, \cdots , i_k \} $ of $\{1, \cdots , m\},$ denote the $k$-th elementary
symmetric function in the variables $u_1, \cdots , u_m.$ Then we get the result as follows.\\

\begin{prop}\label{w4.1} The post quantum blossom of the monomial $M_{k}^
m (t) = t^k$ (considered as a polynomial of
degree $m$) is given by

\begin{equation}\label{e4.2}
m_{k}^m
(u_1, \cdots , u_m; p,q) =
\frac{\phi_{m,k}(u_1,u_2,\cdots,u_m)}
{\phi_{m,k} (p^{m-1},p^{m-2} q, \cdots , q^{m-1})},
\end{equation}

provided that $\phi_{m,k} (p^{m-1},p^{m-2} q, \cdots , q^{m-1}) \neq 0.$
\end{prop}

%

\begin{lem}\label{w4.2}
\begin{equation}\label{e4.3}
\phi_{m,k}(p^{m-1},p^{m-2} q , \cdots, q^{m-1})  = (pq)^{\frac{k(k-1)}{2}}\left[
\begin{array}{c}
m \\
k%
\end{array}%
\right] _{p,q} ~~~~k=0,1,\cdots,m.
\end{equation}
\end{lem}

\textbf{Proof.} Using induction on $ m,$ we get the required result.\\

\begin{cor}\label{w4.3} 
	$ \phi_{m,k} (p^{m-1},p^{m-2} q , \cdots , q^{m-1}) = 0$ if and only if one of the following three conditions is
satisfied:\\
1. $p = 0~~$ and~ $m>1$, $k>1$,~~  $(p=0~ and~2\leq k\leq m)$\\
2. $q = 0~~$ and~ $m > 1$, $k > 1$,~~  $(q=0~ and~ 2\leq k\leq m)$ \\
3. $p = -q$ and~ $m$ is even, $k$ is odd.\\
\end{cor}

\textbf{Proof.} It can be observed that the only real root of a post quantum binomial coefficient can be $p = -q$ because $[m]_{p,q} = \frac{p^m-q^m}{p-q}$ when $p\neq q.$ Condition $1$ and $2$ follows from \ref{e4.3} while Condition 3 follows from the observation that $p=-q$ is a zero of the binomial coefficient $\left[
\begin{array}{c}
n \\
k%
\end{array}%
\right] _{p,q} $ of multiplicity
\begin{equation*}
 \bigg \lfloor \frac {m}{2} \bigg \rfloor -\bigg \lfloor \frac {k}{2}\bigg \rfloor-\bigg \lfloor \frac {m-k}{2}\bigg \rfloor = \left\{
                                                                                                                                       \begin{array}{ll}
                                                                                                                                         1, & \hbox{if m is even and k is odd} \\
                                                                                                                                         0, & \hbox{otherwise.}
                                                                                                                                       \end{array}
                                                                                                                                     \right.
\end{equation*}

Now the existence and uniqueness of the post quantum blossom will be established for all polynomials of degree $m$ and for all real values of $p,q$ that satisfy:\\
\begin{equation}\label{ee4.4}
~~~~~(1)~~~~ q \neq 0 ~~\text{and ~} p \neq 0 ~~\textbf{for all}~~ m>1
\end{equation}
and
\begin{equation}\label{ee4.5}
(2)~~~~ q \neq -p~~ \textbf{for all even}~~ m>1.
\end{equation}

Conditions \ref{ee4.4} and \ref{ee4.5} are now the standard restrictions on the value of $ p,~q.$ \\

From now on, whenever there is a talk of $(p,q)$-blossom or $(p,q)$-Bernstein basis functions or $(p,q)$-B$\acute{e}$zier curves, the standard restrictions stated above will be applicable for the value of $ p,~q$  until it is explicitly mentioned otherwise.\\

\begin{thm}\label{t3.4} \textbf{(Existence and Uniqueness of the post quantum Blossom)}. Corresponding to every polynomial S(t) of at most degree $m,$ there exists a unique symmetric multiaffine function $s(u_1, \cdots, u_m; p,q)$ that reduces to S(t) along the post quantum diagonal. That is, there exists a unique post quantum blossom $s(u_1, \cdots , u_m; p,q)$ for every polynomial S(t) provided that $ p,~q$ satisfies the standard restrictions given by \ref{ee4.4} and \ref{ee4.5}.
\end{thm}

\begin{cor}\label{w4.5}
	 The post quantum blossom of the polynomial $S(t)=\sum \limits_{k=0}^{m} a_k~ t^k $ is
\begin{equation}\label{e4.6}
   s(u_1, \cdots , u_m; p,q) = \sum \limits_{k=0}^{m} a_k \frac {\phi_{m,k}(u_1,u_2, \cdots,u_m)} {(pq)^{\frac{k(k-1)}{2}}\left[
\begin{array}{c}
m \\
k%
\end{array}%
\right]_{p,q}}.
\end{equation}
\end{cor}
In this section, study has been done on the post quantum blossom of a polynomial using the monomial representation. In the next section, investigation will be carried out that how post quantum blossom of polynomial is related to the post quantum Bernstein representation.\\

\section {Post quantum blossoming and Post quantum de Casteljau algorithms}

In the diagrams below, we use the multiplicative notation $u_1 \cdots u_m$ to represent the post quantum blossom value $s(u_1, \cdots , u_m; p,q)$. Though an abuse of notation, this multiplicative notation is highly suggestive. For example, multiplication is commutative and the post quantum blossom is symmetric
\begin{equation*}
    u_1  \cdots u_m = u_{\sigma(1)} \cdots u_{\sigma(m)} \longleftrightarrow s(u_1, \cdots , u_m;p,q) = s(u_{\sigma(1)}, \cdots , u_{\sigma(m)}; p,q).
\end{equation*}

Moreover, multiplication distributes through addition and the post quantum blossom is multiaffine. Thus
$$u = \frac{b-u}{b-a} a + \frac{u-a}{b-a} b$$
implies both
\begin{equation*}
    u_1  \cdots u_m u = \frac{b-u}{b-a} u_1 \cdots  u_m a + \frac{u-a}{b-a} u_1 \cdots  u_m b
\end{equation*}

and
\begin{equation*}
   s(u_1, \cdots , u_m u; p,q) = \frac{b-u}{b-a} s(u_1, \cdots , u_m a; p,q) + \frac{u-a}{b-a} s(u_1, \cdots , u_m b; p,q).
\end{equation*}

\begin{figure*}[htb!]
\begin {center}
\begin {tikzpicture}[-latex ,auto ,node distance =2 cm and 2cm ,on grid ,
semithick ,
state/.style ={ circle ,top color =white , bottom color = processblue!20 ,
draw,processblue , text=blue , minimum width =.5 cm}]
\node[state] (A)
{$tp^2~tpq~tq^2$};
\node[state] (B) [below left=of A] {$0~tpq~tq^2$};
\node[state] (C) [below right =of A] {$p^2~tpq~tq^2$};
\node[state] (D) [below left=of B] {$0~0~tq^2$};
\node[state] (E) [below right=of B] {$0~p^2~tq^2$};
\node[state] (F) [below right=of C] {$p^2~pq~tq^2$};
\node[state] (G) [below left=of D] {$0~0~0$};
\node[state] (H) [below right=of D] {$0~0~p^2$};
\node[state] (I) [below right=of E] {$0~p^2~pq$};
\node[state] (J) [below right=of F] {$p^2~pq~q^2$};

\path (B) edge [bend right =0] node[left = 0.05 cm] {$1-t$} (A);
\path (C) edge [bend right = 0] node[right =0.05 cm] {$t$} (A);
\path (D) edge [bend left =0] node[left =0.05 cm] {$p-tq$} (B);
\path (E) edge [bend left =0] node[left =0.05 cm] {$tq$} (B);
\path (E) edge [bend left =0] node[left =0.05 cm] {$p-pt$} (C);
\path (F) edge [bend right = 0] node[right =0.05 cm] {$pt$} (C);
\path (G) edge [bend right = 0] node[left =0.05 cm] {$p^{2}-tq^2$} (D);
\path (H) edge [bend right = 0] node[left =0.05 cm] {$tq^2$} (D);
\path (H) edge [bend right = 0] node[left =0.05 cm] {$p^{2}-tpq$} (E);
\path (I) edge [bend right = 0] node[left =0.05 cm] {$ptq$} (E);
\path (I) edge [bend right = 0] node[left =0.05 cm] {$p^2-p^2t$} (F);
\path (J) edge [bend right = 0] node[right =0.05 cm] {$p^2t$} (F);

\end{tikzpicture}
\end{center}
\caption{`Computing $s( p^{m-1}t,p^{m-2}t q ,  \cdots ,tq^{m-1}; p,q)=S(t)$ recursively from the initial post quantum blossom values $s(0, \cdots , 0, p^{m-1},p^{m-2} q ,  \cdots ,p^{m-k}q^{k-1}; p,q),~~  k=0,1,\cdots,m.$' .}\label{f3}
 We illustrate the cubic case relative to Figure 1
\end{figure*}
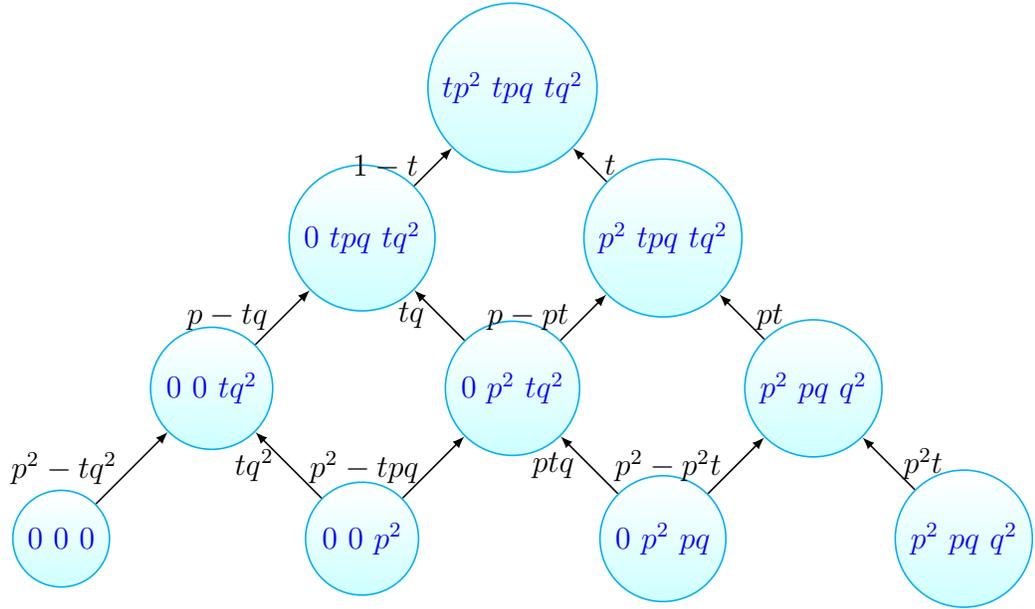

The diagram represents the symmetry and multiaffinity character while at same time also make the case for multiplication and post quantum blossoming both. Therefore this multiplicative representation for the post quantum blossom seems to be natural. Due to similarity between multiplication and post quantum blossoming, identities corresponding to multiplication expect to give an analogous identities for the post quantum blossom.\\

Using this multiplicative notation, Figure 3 shows (for $m = 3$) how to compute an arbitrary value of $s(p^{m-1}t,p^{m-2} q t, \cdots , q^{m-1}t;p,q) = S(t)$ recursively from the initial post quantum blossom values $s(0, \cdots , 0, p^{m-1},p^{m-2} q , \cdots , p^{m-k}q^{k-1}; p,q)$, with exactly $m - k$ blossom values set to $0$ for $k = 0,1, \cdots , m$, by applying the multiaffine and symmetry properties at each node.\\

Now compare the post quantum blossoming algorithm in Figure 3 to the de-Casteljau algorithm in Figure 1 for post quantum B$\acute{e}$zier curves. For arbitrary $m,$ Figures 1 and 3 are similar, and Figure 3 is this de Casteljau algorithm for $s(p^{m-1}t,p^{m-2} q t, \cdots , q^{m-1}t; p,q) = S(t)$ with control points $s(0, \cdots , 0, p^{m-1},p^{m-2} q , \cdots ,p^{m-k} q^{k-1}; p,q)$, $k = 0,1, \cdots , m$. In next three theorems, certain observations having some important consequences for arbitrary values of the degree $m$ has been done. standard restrictions given by \ref{ee4.4} and \ref{ee4.5} on the value of $p,q$ will be applicable until mentioned otherwise.\\

 \begin{thm}\label{t4.1}
  Any polynomial represents post quantum B$\acute{e}$zier Curve. In other words, let S(t) be a polynomial of degree m with post quantum blossom $s(u_1, \cdots , u_m; p,q)$. Then using de Casteljau algorithm \ref{ee2.3}, S(t) can be generated by control points $P_k = s(0, \cdots , 0, p^{m-1},p^{m-2} q , \cdots , p^{m-k} q^{k-1}; p,q)$,  $k = 0,1, \cdots , m.$

\end{thm}


\begin{cor}
 On interval $[0, 1]$, $m$ degree post quantum Bernstein basis functions form the basis for $m$ degree polynomial, except when $(p=0~for~m>1)$ ~and ~$(q = -p ~ for~ even~ m)$. 
\end{cor}

\textbf{Proof.} 
Result can be drawn directly from Theorem \ref{t4.1} when $p$ and $q$ satisfies the standard restrictions given by \ref{ee4.4} and \ref{ee4.5}. Further, when $q=0\neq p$ this result can be obtained explicitly by using formula for basis in \ref{ee4}, since\\
$B_i^m(t;p,0)=t^i-t^{i+1}$,~~  $i=0,1, \cdots ,m-1$\\
$B_m^m(t;p,0)=t^m.$

\begin{cor}
 On interval $[0, 1]$, post quantum B$\acute{e}$zier curve's control points are unique.\\
\end{cor}

%
%
%

\begin{thm}\label{t4.4} \textbf{(Dual Functional Property of the post quantum Blossom}). Let $S(t)$ be a post quantum B$\acute{e}$zier curve of degree m and let $s(u_1, \cdots , u_m; p,q)$ be the post quantum blossom of S(t). Then the post quantum B$\acute{e}$zier control points of S(t) are given by\\
\begin{equation}\label{y4.3}
P_k= s(0, \cdots , 0, p^{m-1},p^{m-2} q , \cdots , p^{m-k} q^{k-1} ; p,q), ~~~ k = 0,1, \cdots , m.
\end{equation}
\end{thm}

\begin{thm}\label{t4.5} Let $S(t)=\sum\limits_{i=0}^{m}$ $P_i ~B_i^m(t;p,q)$ be a post quantum B$\acute{e}$zier curve of degree $m$ with post quantum Blossom $s(u_1, \cdots , u_m; p, q).$ Define recursively a set of multiaffine functions by setting $Q_i^0=P_i,~i=0,..., m$ and

\begin{equation}\label{e4.5}
Q_i^{k+1} (u_1, \cdots , u_{k+1})=(1-u_{k+1}~ p^{i}~q^{-i})Q_i^{k} (u_1, \cdots , u_{k})+ u_{k+1}~  p^{i}~q^{-i} Q_{i+1}^{k} (u_1, \cdots , u_{k})
\end{equation}

$i=0,1, \cdots ,m-k-1$ and $k=0,1,\cdots,m-1.$ Then

$$ Q_i^{k} (u_1, \cdots , u_{k})= s(0, \cdots , 0, p^{m-1},p^{m-2} q , \cdots ,  p^{m-i} q^{i-1},u_1, \cdots , u_k; p, q)$$

$i=0,1, \cdots ,m-k$ and $k=0,1, \cdots ,m.$ \\\\
In particular,

$ Q_0^{i} (u_1, \cdots , u_{m})= s(u_1, \cdots , u_m; p, q)$

\end{thm}

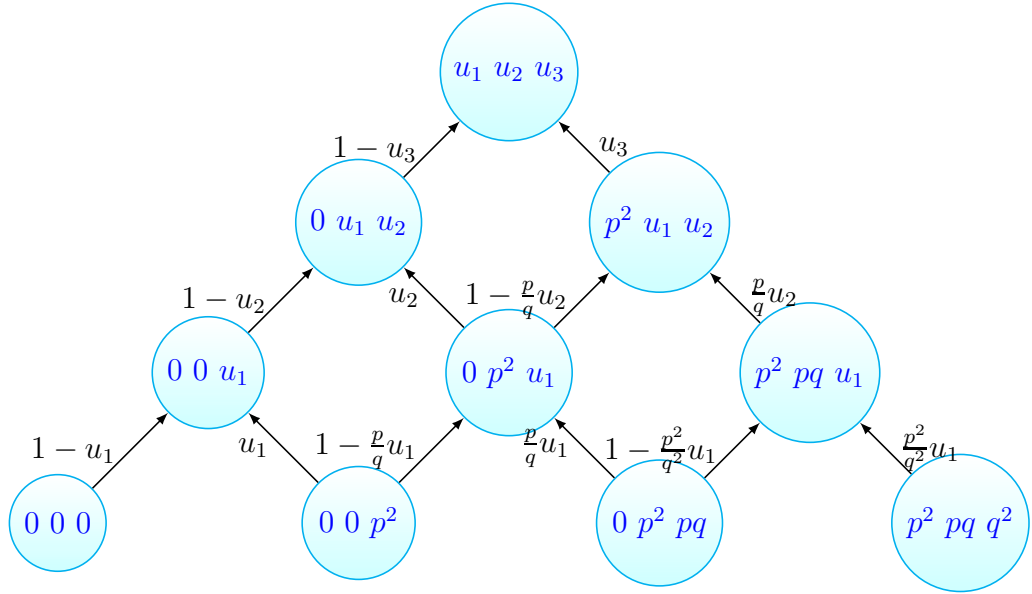
\begin{figure*}[htb!]
\begin {center}
\begin {tikzpicture}[-latex ,auto ,node distance =2 cm and 2cm ,on grid ,
semithick ,
state/.style ={ circle ,top color =white , bottom color = processblue!20 ,
draw,processblue , text=blue , minimum width =.5 cm}]
\node[state] (A)
{$u_1~u_2~u_3$};
\node[state] (B) [below left=of A] {$0~u_1~u_2$};
\node[state] (C) [below right =of A] {$p^2~u_1~u_2$};
\node[state] (D) [below left=of B] {$0~0~u_1$};
\node[state] (E) [below right=of B] {$0~p^2~u_1$};
\node[state] (F) [below right=of C] {$p^2~pq~u_1$};
\node[state] (G) [below left=of D] {$0~0~0$};
\node[state] (H) [below right=of D] {$0~0~p^2$};
\node[state] (I) [below right=of E] {$0~p^2~pq$};
\node[state] (J) [below right=of F] {$p^2~pq~q^2$};

\path (B) edge [bend right =0] node[left = 0.05 cm] {$1-u_3$} (A);
\path (C) edge [bend right = 0] node[right =0.05 cm] {$u_3$} (A);
\path (D) edge [bend left =0] node[left =0.05 cm] {$1-u_2$} (B);
\path (E) edge [bend left =0] node[left =0.05 cm] {$u_2$} (B);
\path (E) edge [bend left =0] node[left =0.05 cm] {$1-\frac{p}{q}u_2$} (C);
\path (F) edge [bend right = 0] node[right =0.05 cm] {$\frac{p}{q}u_2$} (C);
\path (G) edge [bend right = 0] node[left =0.05 cm] {$1-u_1$} (D);
\path (H) edge [bend right = 0] node[left =0.05 cm] {$u_1$} (D);
\path (H) edge [bend right = 0] node[left =0.05 cm] {$1-\frac{p}{q}u_1$} (E);
\path (I) edge [bend right = 0] node[left =0.05 cm] {$\frac{p}{q}u_1$} (E);
\path (I) edge [bend right = 0] node[left =0.05 cm] {$1-\frac{p^2}{q^2}u_1$} (F);
\path (J) edge [bend right = 0] node[right =0.05 cm] {$\frac{p^2}{q^2}u_1$} (F);

\end{tikzpicture}
\end{center}
\caption{`Recursive evaluation algorithm for the post quantum blossom of a Cubic post quantum B$\acute{e}$zier curve'}\label{f4}
\end{figure*}

\textbf{Proof.} By the dual functional property, $$ Q_i^{0} (u_1, \cdots , u_{m})= s(0, \cdots , 0, p^{m-1},p^{m-2} q , \cdots ,  p^{m-k} q^{k-1}; p, q)~,~~~~i = 0,1, \cdots , m. $$ 

By applying induction on $k,$ rest of the proof can be easily done.  The case $m = 3 $ is illustrated by Figure 4.


\begin{thm}\label{t4.6}  Let $S(t)=\sum\limits_{i=0}^{m}$ $P_i ~B_i^m(t;p,q)$ be a post quantum B$\acute{e}$zier curve of degree $m$ with post quantum Blossom $s(u_1, \cdots , u_m; p, q).$ There are $m!$ affine invariant, recursive evaluation algorithms for S(t) defined
as follows: Let $\sigma$ be a permutation of $\{1,2, \cdots , m\} $ and let $P_{i}^0 (t) = P_i$ , $i = 0,1, \cdots , m.$ Define

\begin{equation}\label{e4.7}
P_{i}^{k+1}(t;p,q)= (p^{\sigma (k+1)-1}-t~p^{i}~q^{\sigma (k+1)-1-i})P_{i}^{k}(t;p,q) + p^{i}~q^{\sigma (k+1)-1-i}~t~ P_{i+1}^{k}(t;p,q)
\end{equation}

$i=0,1, \cdots  ,m-k-1$ and $k=0,1,  \cdots ,m-1.$  ~Then

\begin{equation}\label{e4.8}
P_{i}^{k}(t)=s(0, \cdots  , 0, p^{m-1},p^{m-2} q , \cdots  , p^{m-i}q^{i-1}, t p^{\sigma (1)-1}~q^{m-\sigma (1)}, \cdots  ,t p^{\sigma (k)-1}~q^{m-\sigma (k)}; p, q)
\end{equation}

$i=0,1,\cdots,m-k$ and $k=0,1,\cdots ,m.$ \\

In particular

\begin{equation}\label{e4.9}
P_{0}^{m}(t)=s( t p^{\sigma (1)-1}~q^{m-\sigma (1)}, \cdots  , t p^{\sigma (m)-1}~q^{m-\sigma (m)}; p, q) =S(t).
\end{equation}

\end{thm}


\section{ Identities for Bernstein basis functions based on post quantum blossoming}

Three identities have been derived for the post quantum Bernstein basis functions in this section. Each of these identities  can be expressed into standard quantum Bernstein basis functions after putting $p=1$. Standard restrictions on $p$ and $q $ given by \ref{ee4.4} and \ref{ee4.5}. Starting from new variant of Marsden's identity.\\

\begin{prop} \textbf{~(~Marsden's~ Identity~)}
\begin{equation}\label{e5.1}
\prod\limits_{i=1}^{m}(p^{i-1}x-q^{i-1}t)= \sum\limits_{j=0}^{m} \frac{(-1)^{j}~ p^{\frac{(m-j)(m-j-1)}{2}}~~ q^{\frac{j(j-1)}{2}}~~B_{m-j}^m(x;\frac{1}{p},\frac{1}{q})~~B_{j}^m(t;p,q)} {\left[\begin{array}{c}
m \\
j%
\end{array}%
\right]_{\frac{1}{p},\frac{1}{q}}}.
\end{equation}
\end{prop}


Monomials can also be expressed in terms of the post quantum Bernstein basis functions.

\begin{prop} \textbf{~(~Monomial ~Representation~)}
\begin{equation}\label{5.2}
  t^i=\sum\limits_{k=i}^{n}~~p^{i(m-k)}~~\frac{\left[\begin{array}{c}
k \\
i%
\end{array}%
\right] _{p,q} }{\left[\begin{array}{c}
m \\
i%
\end{array}%
\right] _{p,q} }~~ B_{k}^m(t;p,q)~~~~i=0,1,\cdots,m.
\end{equation}
\end{prop}

Reparametrization formula for post quantum Bernstein basis functions is last identity of this section which has its use in subdivision algorithms for B$\acute{e}$zier curves. Before proceeding to proof this change of basis formula, first there is a need to know a lemma.\\

\begin{lem}\label{w6.3} 
	Let $b_i^m(u_1, \cdots , u_m; p,q)$ denote the $(p,q)$-blossom of $B_i^m (t; p,q),~~ i = 0, 1,\cdots , m.$ Then
\begin{equation*}
 b_{i}^{m}(u_1, \cdots , u_{m-1}, 0; p,q) = b_{i}^{m-1}(u_1, \cdots , u_{m-1}; p,q), ~~i = 0,1, \cdots , m - 1.
\end{equation*}
\end{lem}

\begin{prop} \textbf{(Reparametrization Formula)}
\begin{equation*}
  B_k^m(rt; p,q) =\sum_{i=k}^{m} B_k^i(r ; p,q)B_i^m(t; p,q).
\end{equation*}
\end{prop}




%
}
%

\end{document}